\documentclass[final,leqno]{siamltexmm2}
\usepackage{amsmath,amssymb}

\usepackage{graphicx}
\usepackage{dcolumn}
\usepackage{bm}
\usepackage{epstopdf,overpic,color,rotating}
\usepackage{overpic}
\usepackage{gensymb}
\usepackage[table]{xcolor}
\usepackage[utf8]{inputenc}
\usepackage{textcomp,marvosym}

\graphicspath{{figures/}}

\newcommand*\samethanks[1][\value{footnote}]{\footnotemark[#1]}

\title{Data-Driven Forecast of Dengue Outbreaks in Brazil: A Critical Assessment of 
Climate Conditions for Different Capitals.}

\author{Lucas M. Stolerman$^\ddagger$\thanks{Instituto Nacional de Matem\'atica Pura e Aplicada (IMPA), Rio de Janeiro, Brazil.  $^\ddagger$ (\email{Lucasms@impa.br}). Questions, comments, or corrections to this document may be directed to that email address. }
 \and Pedro D. Maia\thanks{Department of Applied Mathematics, University of Washington, Seattle, WA. 98195-2420.  (\email{pmaia@uw.edu} , \email{kutz@uw.edu})
}
\and J. Nathan Kutz\samethanks[2] }

\begin{document}
\maketitle
\newcommand{\slugmaster}{%
\slugger{siads}{xxxx}{xx}{x}{ }}

\begin{abstract}
Local climate conditions play a major role in the development of the mosquito 
population responsible for transmitting Dengue Fever. Since the {\em Aedes Aegypti} mosquito is also a primary vector for the recent Zika and Chikungunya epidemics across the Americas, a detailed
monitoring of periods with favorable climate conditions for mosquito profusion may improve 
the timing of vector-control efforts and other urgent public health strategies. 
We apply dimensionality reduction techniques and machine-learning algorithms to climate 
time series data and analyze their connection to the occurrence of Dengue outbreaks for seven major cities in Brazil. 
Specifically, we have identified two key variables and a period during the annual cycle that are highly predictive of
epidemic outbreaks.  The key variables are the frequency of precipitation and temperature during an approximately two month window of the winter season preceding the outbreak.  Thus simple climate signatures may be influencing Dengue outbreaks even months before their occurrence. 
Some of the more challenging datasets required usage of compressive-sensing procedures to estimate missing entries for temperature and precipitation records.  Our results indicate that each Brazilian capital considered has a unique frequency of precipitation and temperature signature in the winter preceding a Dengue outbreak.  Such climate contributions on vector populations are key factors 
in dengue dynamics which could lead to more accurate prediction models and early warning systems. 
Finally, we show that critical temperature and precipitation signatures may vary significantly from city to 
city, suggesting that the interplay between climate variables and dengue outbreaks is more complex than 
generally appreciated.
\end{abstract}

\section*{Author Summary}
The high number of dengue, chikungunya and zika viral infections in Brazil were a
major cause for concern during Rio de Janeiro's 2016 Olympic games. News reports on microcephalic
newborns in the poor northern regions of the country alarmed the world
and urged public health authorities to take action. Since all such viruses are transmitted by the same
disease vector, the moquito \emph{Aedes Aegypti}, a better understanding of which climate conditions favor their proliferation is crucial for vector-control strategies. We analyzed climate data from 7 Brazilian state capitals and found that climate effects often occurring months before the outbreaks may be of critical importance for prediction. Thus, the early incidence of key temperature and precipitation signatures -- that vary from city to city -- should urge municipal health authorities to anticipate early interventions and ensure adequate responses to the local climate conditions.
%
%

\newpage
\section*{Introduction}
Dengue Fever is a tropical mosquito-borne viral disease present in more than 110 countries
and a current threat to half of the world population ~\cite{Ranjit2011,Gubler1998,who,Bhatt}. The DENV virus -- and the more perilous Chikungunya and Zika virus -- are primarily transmitted to humans through 
infected  \emph{Aedes Aegypti} mosquitoes, which were the subject of much debate during the 2016 Olympic
Games in Rio de Janeiro. This main disease vector is well adapted to urban environments, which allow viruses to spread easily through cities.  Still, regional climate conditions play a critical role in the development of epidemic outbreaks in major urban centers. In this work we analyze temperature and precipitation time series data for Brazilian state capitals and determine critical periods and seasons in which these climate variables might favor the mosquito development cycle and therefore the occurrence of Dengue outbreaks.

The first cases of Dengue in Brazil date from the end of the $19^{th}$ century, and despite the elimination of 
the \emph{Aedes Aegypti} in 1955, the mosquito was reintroduced in the country in the 70s. A historically important 
outbreak occurred in 1981 in Boa Vista, in the state of Roraima, following several outbreaks in Central America involving 
the DENV-1 and DENV-4 serotypes ~\cite{Figueredo, Fares}. Since then, Dengue has become one of the major public 
health problems in Brazil, with several epidemics reported yearly across the country. While Dengue symptoms are
 usually limited to fever and muscle/joint pain, some develop more severe forms of the disease such as hemorrhagic fever or shock 
syndrome. The epidemics were aggravated with the latest Zika and Chikungunya developments. In fact, 
$91,387$ thousand cases of Zika and 39,017 thousand cases of Chikungunya were reported in 2016 from 
February to April alone ~\cite{Saude}, which caught the world's attention just in time for the Olympic Games in Rio de Janeiro. Until recently, Brazilian authorities limited their actions to vector control measures, 
but a first-generation vaccine may represent a turning point for stopping these epidemics ~ \cite{who_vaccine,Rothman,Pitisuttithum}. 

The proliferation of \emph{Aedes aegypti} and the sustained transmission of Dengue are influenced by a complex, interplay of multi-scale factors such as the circulation of different serotypes ~\cite{Rabaa,Raghwani},
the commuting of infected and susceptible humans within a city 
~\cite{manbites,stoddard,stolerman},  and the 
population size of the mosquitoes. There is also a growing body of evidence showing that local climate conditions
such as temperature and precipitation may highly influence the development of the mosquitoes 
throughout the different stages of their life cycle ~\cite{Watts,Yang,Honorio,Hopp}. Complicating our understanding
is the fact that several regions experienced a nontrivial alternation between periods with and without epidemic outbreaks 
over the past years, suggesting that the specific critical climate conditions that propitiate the transmission of the disease is heterogeneous and 
still poorly understood ~ \cite{Adde,Hii, Descloux,Buczak2012,Buczak2014}. 

\begin{figure}[t]
\includegraphics[width=1\textwidth]{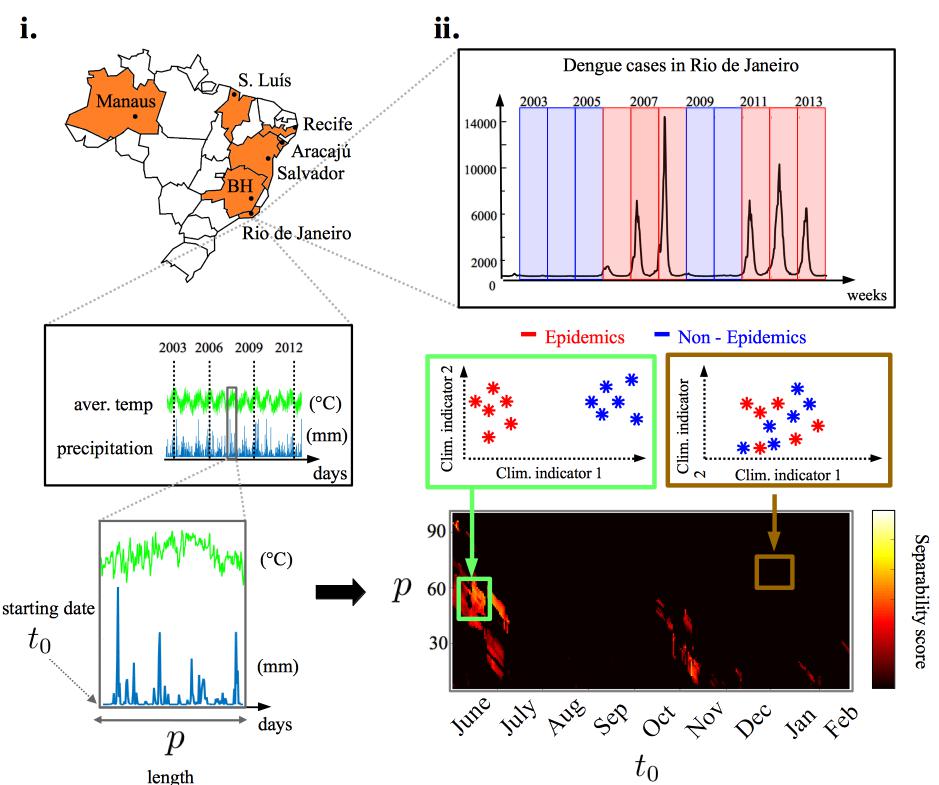}
\caption{ {\bf Schematic Overview.}
We analyze time series data for climate variables from seven Brazilian state capitals (Aracaj\'{u}, Belo Horizonte, Manaus, Recife, Rio de 
Janeiro, Salvador and S\~{a}o Lu\'{i}s) and their connection to Dengue outbreaks. {\bf(i)} Illustrative example showing data from Rio de Janeiro. 
Two parameters define the epochs in which climate conditions are considered: the starting date $t_0$ (month/day) and period length $p$ (days). {\bf(ii)} By applying machine-learning algorithms to historical data we locate periods along the year 
where the separability between epidemic and non-epidemic climate is higher. Keeping track of signature differences at key epochs, that 
vary from capital to capital, may significantly improve Dengue outbreak forecasting in the upcoming years. }
\label{Schematic_Overview}
\end{figure}

In this work we analyze climate and epidemiological data from seven major Brazilian cities that in the recent past 
had years with and without Dengue outbreaks in order to identify critical climate signatures that may have contributed 
to the epidemic outcomes.  Figure ~\ref{Schematic_Overview} is a schematic overview of the work presented. We estimate the influence of 
climate conditions in different epochs preceding epidemic periods using two data-driven methodologies; the first one is 
based of the singular value decomposition and exploits the low dimensional structures present in the climate time series 
~\cite{Golub,Kutz}, and the second one is based of machine learning algorithms for clustering and classification~\cite{murphy,bishop} such as Support Vector Machines (SVM) 
applied to climate variables that are key to the life cycle of the mosquito ~\cite{Vapnik,Burges}. A crucial step in our methodology 
includes the usage of compressed sensing to recover missing data~\cite{Kutz,cs1,cs2,cs3,cs4} -- in a plausible and in a $\mathcal{L}^{1}$-optimal way -- 
from climate recordings by the National Institute of Meteorology (INMET) ~\cite{INMET}. 
This allow us to explore the link between climate and Dengue in 
the following  major Brazilian cities : Aracajú, Belo Horizonte, Manaus, Recife, Rio de Janeiro, Salvador and São Luís. 
For each city, we highlight epochs that are critical for both methodologies. Surprisingly, there is a strong correlation between 
dengue epidemics and favorable climate conditions during winter and spring. This long-term influence is important evidence 
that the interplay between climate, mosquito populations and dengue outbreaks are extremely complex.
The insights of this work may help taylor public health policies for each different city by increasing vector control measures 
during neglected critical epochs and ultimately improving the forecasting of Dengue outbreaks -- which would allow the public 
health system to make earlier logistic preparations to better accommodate a large number of patients, or alternatively, mosquito eradication programs can be enacted during the winter and spring months that are known to be associated with epidemic outbreaks.

The outline of the paper is as follows. In the \textbf{Methods} section,  we describe both epidemiological and climate datasets, our techniques for data completion and other details of our analysis.  In the \textbf{Results} section we present our findings for all seven Brazilian capitals, emphasizing epochs that are critical for all methods. In the \textbf{Discussion} section we summarize the most important seasons for dengue epidemics in each city, highlighting the long-term impact of climate. We also discuss the limitations of this work,  its potential impact for improving early warning systems, and the usage of our methods as a modest outbreak prediction tool. 
%
%
%
\section*{Methods}
%
%
%
\subsubsection*{Description of epidemiological and climate datasets}

All epidemiological data utilized in this work were taken from the publicly available 
datasets of the Brazilian Notifiable Diseases Information System (SINAN, ~\cite{SINAN}). 
This includes the total number of Dengue cases per year (from 2002 to 2012) for all Brazilian 
state capitals. We also include data made available for Rio de Janeiro by the city's hall health 
department for 2013~\cite{Riodengue}. A year is conventionally classified as an \emph{epidemic} year 
for a given city if the incidence of Dengue is above 100 cases (per 100,000 inhabitants) and 
classified as a \emph{non-epidemic} year otherwise. In order to find critical climate signatures that may 
have contributed to the epidemic outcomes, we restrict ourselves to seven state capitals that 
displayed at least 3 epidemic years and 3 non-epidemic years in the recent past. This allowed 
us to investigate the correlation between distinct climate conditions and the complicated alternations 
between years with and without epidemic outbreaks over time. The climate data utilized in this 
work was obtained from the National Institute of Meteorology (INMET)  and included 
time series for the average temperature and precipitation for the state capitals Aracajú, Belo 
Horizonte, Manaus, Recife, Salvador, and São Luís (from 1/1/2001 to 12/31/2012) and for 
Rio de Janeiro (from 1/1/2002 to 12/31/2013).

\subsubsection*{Completing missing climate data via compressive sensing}
\begin{figure}[!t]
\includegraphics[width= 1\textwidth]{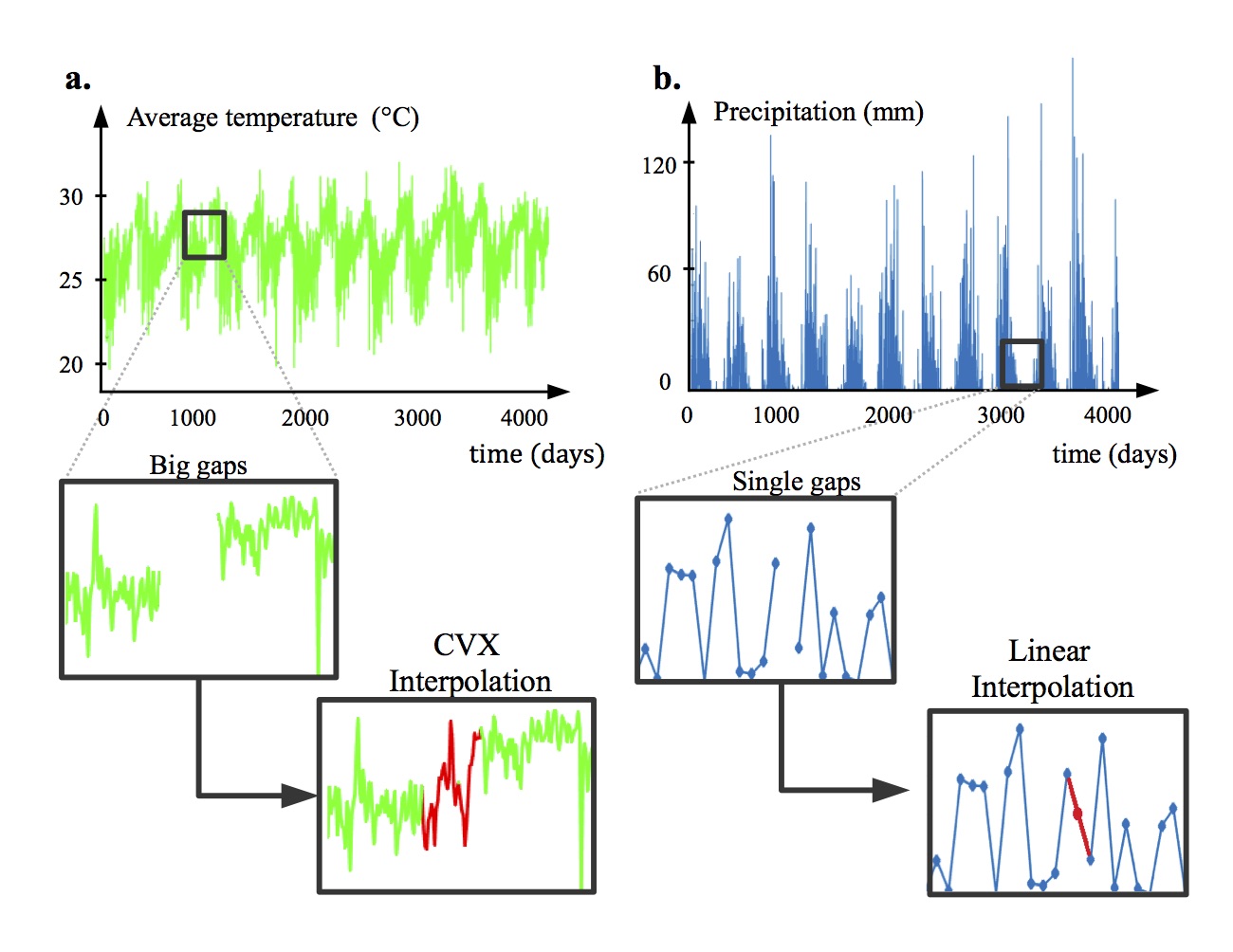}
\caption{{\bf Completing missing data.} The daily measurements of climate variables for Brazilian state capitals 
from the National Institute of Meteorology  (\href{http://www.inmet.gov.br/portal/}{INMET}) contain episodical gaps. 
\textbf{a.} We reconstruct larger portions of lacking data with compressed sensing ($\mathcal{L}^{1}$-convex 
optimization routines ). \textbf{b.} Data values at minor holes were estimated by simpler
interpolation protocols. Capitals with intractable missing portions of data were not considered (see appendix)
for more details.}
\label{fig1}
\end{figure}

The time series of our selected climate dataset contain episodical gaps on days where 
variables (temperature and precipitation) were not recorded. To fill in the missing data gaps, we employ two 
different methods: compressive sensing~\cite{Kutz,cs1,cs2,cs3,cs4} and interpolation (see Fig.~\ref{fig1} for illustrative examples). For temperature time series data with 2 or more consecutive missing recordings, we 
use a recently developed compressive sensing method based upon $\mathcal{L}^{1}$-convex optimization
for approximating the missing data~\cite{Kutz,cs1,cs2,cs3,cs4}.  
The compressive sensing method attempts to reconstruct a signal from a sparse, subsampling
of the time series data.  In this case, the sparse subsampling occurs from the fact that we have missing
data.  The signal reconstruction problem is nothing more than a large underdetermined system of linear equations.  To be more precise, consider the conversion of a time series data to the frequency domain via the discrete cosine transform (DCT) 
\begin{equation}
  {\bf \psi} {\bf c}={\bf f}
\end{equation}
where ${\bf f}$ is the signal vector in the time domain and ${\bf c}$
are the cosine transform coefficients representing the
signal in the DCT domain.  The matrix
${\bf \psi}$ represents the DCT transform itself.  The
key observation is that most of the coefficients of 
the vector ${\bf c}$ are zero, i.e. the time series is sparse in the Fourier domain.  
Note that the matrix ${\bf \psi}$ is
of size $n\times n$ while ${\bf f}$ and ${\bf c}$ are $n\times 1$
vectors.  The choice of basis functions is critical in carrying out
the compressed sensing protocol.  In particular, the signal
must be sparse in the chosen basis.  For the example here
of a cosine basis, the signal is clearly sparse, allowing us
to accurately reconstruct the signal using sparse sampling.
The idea is to now sample the signal randomly (and sparsely)
so that
\begin{equation}
     {\bf b}= {\bf \phi} {\bf f}
\end{equation}
where ${\bf b}$ is a few ($m$) random samples of the
original signal ${\bf f}$ (ideally $m\ll n$).  
Thus ${\bf \phi}$ is a subset
of randomly permuted rows of the identity operator.
More complicated sampling can be performed, but this
is a simple example that will illustrate all the key features.
Note here that ${\bf b}$ is an $m\times 1$ vector while
the matrix ${\bf \phi}$ is of size $m\times n$.

Approximate signal reconstruction can then be performed
by solving the linear system
\begin{equation}
  {\bf A}{\bf x}={\bf b}
\end{equation}
where ${\bf b}$ is an $m\times 1$ vector, ${\bf x}$ 
is $n\times 1$ vector and
\begin{equation}
  {\bf A} = {\bf \phi}{\bf \psi}
\end{equation}
is a matrix of size $m\times n$.  Here the ${\bf x}$ is
the sparse approximation to the full DCT coefficient
vector.  Thus for $m\ll n$, the
resulting linear algebra problem is highly underdetermined.  The idea is then to solve the
underdetermined system using an appropriate norm constraint that
best reconstructs the original signal, i.e. the sparsity promoting $\mathcal{L}^{1}$ is highly appropriate.  The signal reconstruction
is performed by using
\begin{equation}
   {\bf f}\approx{\bf \psi} {\bf x} \, .
   \label{eq:cs_reconstruct}
\end{equation}
If the original signal had exactly $m$ non-zero coefficients,
the reconstruction could be made exact (See Ref.~\cite{Kutz}, Ch.~18).

We applied this technique specifically to the climate series of Rio de Janeiro, Salvador and São Luís. 
For the other capitals, we just linearly interpolate the time series whenever a single daily 
recording is missing. We note that there were intractable large gaps for the INMET 
precipitation series for Rio de Janeiro, which forced us to use alternative data sources made available 
by the city's alert system of rain events ~\cite{Alertario}. See the SI tables for details.

\subsubsection*{Defining periods of critical climate conditions for Dengue}
In what follows, we investigate the influence of climate conditions 
on Dengue outbreaks at different periods along the yearly cycle. We 
let \emph{$(t_0,p)$} denote a sampling period of $p$ days starting at the date $t_0$. 
Then, for a fixed period, we evaluate a score quantifying the discrepancy 
between climate conditions in epidemic years and non-epidemic years. 
See Fig. ~\ref{Schematic_Overview} for an illustrative example using data 
from the city of Rio de Janeiro; we postulate that periods with high climate 
\textit{separability} between epidemic years (in red) and non-epidemic years 
(in blue) might be of critical importance to the cycle of the urban mosquito population 
and consequently, to the occurrence of dengue outbreaks in the following year.  
We calculate the separability score of a period using two different methodologies:
the first is based of the Singular Value Decomposition (SVD)~\cite{Kutz} and a low dimensional 
representation of the climate data, while the second is based of a machine learning
algorithm know as support vector machine (SVM)~\cite{murphy,bishop}. In both cases, the methods 
highlight potentially critical periods for the occurrence of Dengue. Finally, 
since Dengue outbreaks in Brazil typically take place between March--May 
in a given year, we limit the range of  \emph{$(t_0,p)$} from June (of the previous year)
to May. In Fig. ~\ref{Schematic_Overview}, we note that there are critical periods
in the winter (green box with $t_0$ in June) that may be critical for the occurrence 
of dengue.
%
%
%
%
%
%
%
\subsubsection*{Separability scores from SVD methodology}
%
%
%
\begin{figure}[t]
\includegraphics[width=1\textwidth]
{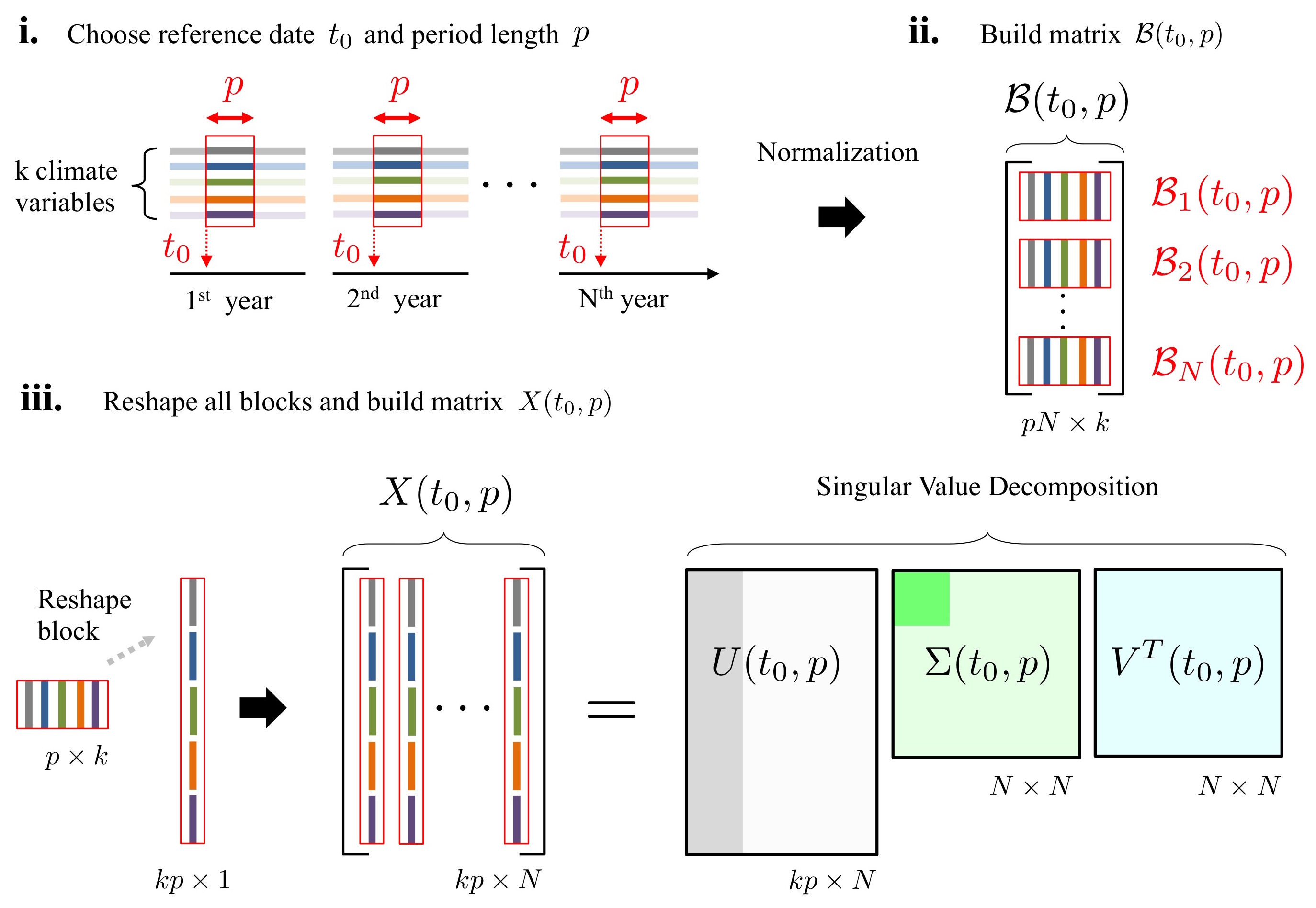}
\caption{{\bf Outline of SVD methodology: 
Data matrix setup.} \textbf{(i)} 
We select climate data with the same starting date 
$t_{0}$ and length $p$ across the years ($1,2 \hdots ,N$).  \textbf{(ii)} After normalizing each climate variable over the years, 
we store them in block matrices  $\mathcal{B}_j(t_{0},p)$, which in turn, are stacked in a matrix $\mathcal{B}(t_{0},p)$. 
\textbf{(iii)} Reshape $\mathcal{B}$ into ${\bf X}$, where different 
columns correspond to climate information collected at $(t_{0},p)$ in different years. 
The SVD of ${\bf X}$ provides a low-dimensional representation of the internal structure of the data from its most informative (correlated) viewpoint. Our goal is to, based in the historic data, determine specific epochs of the year in which the separability between epidemic and non-epidemic climate is higher.}
\label{fig3}
\end{figure}
\begin{figure}[!t]
\centering
\includegraphics[width=.9\textwidth]{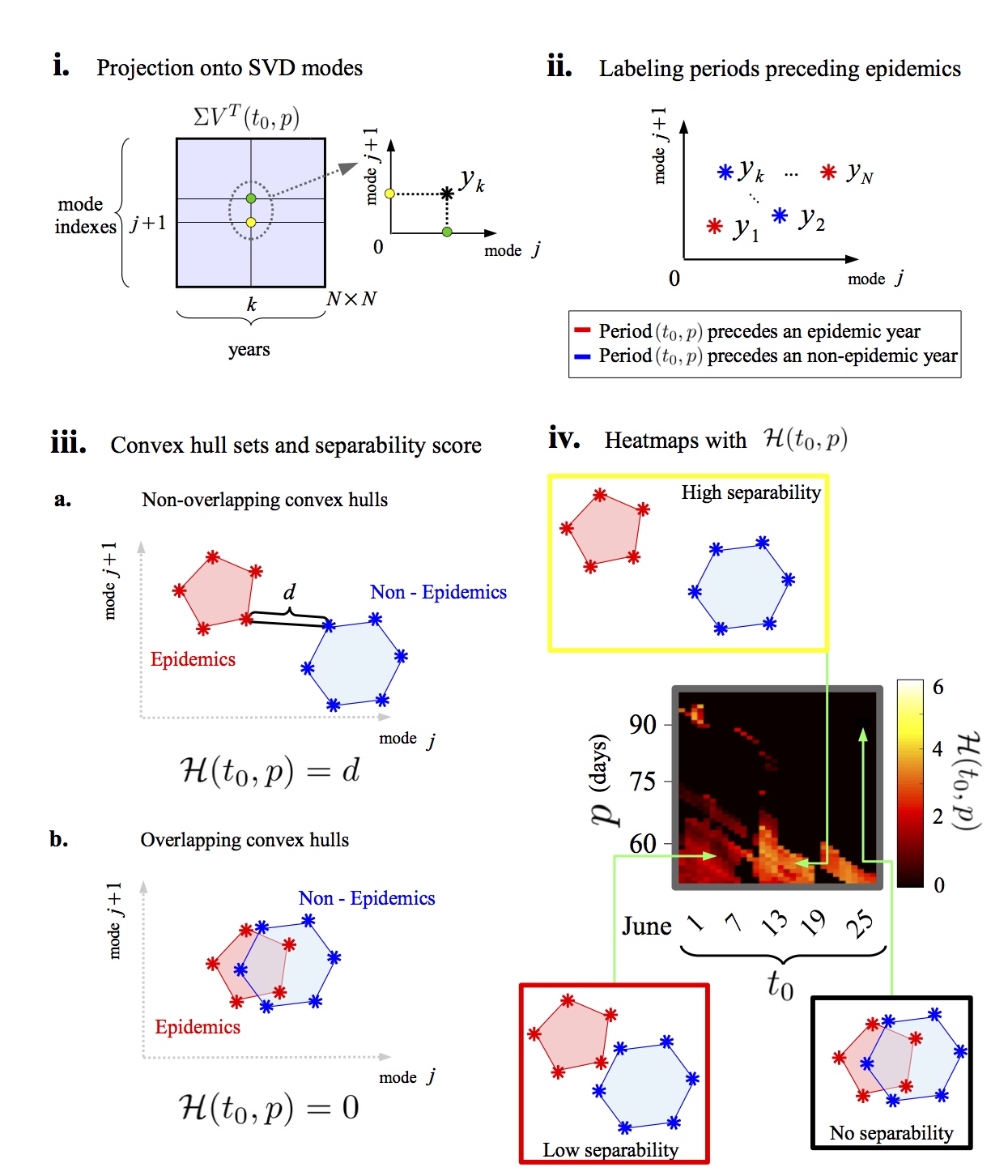}
\caption{{\bf   Outline of SVD methodology: 
Convex Hull analysis.}
\textbf{(i)} The projection's component of the $k$-th column of $X$ onto the $j$-th mode is the $(j,k)$-element of the matrix $\Sigma V^{T}$. We plot the projection for each year $l$ $(l=1,2,\cdots,N)$ in the plane spanned by modes  $j$ and $j+1$.
\textbf{(ii)} For each year we color the projections according to epidemic or non-epidemic year criteria. We choose red If the $(t_0,p)$ interval preceded a DF outbreak and blue if it doesn't.
\textbf{(iii)} We compute the convex hulls for the epidemic and non-epidemic projections set. \textbf{(a)} If there is no overlapping between the hulls, we calculate the minimum distance between two vertices and set  $\mathcal{H}(t_0,p) = d$. \textbf{(b)}  $\mathcal{H}(t_0,p) = 0$ in the case of overlapping hulls. 
\textbf{(iv) } The SVD separability score $\mathcal{H}$ can be obtained for a range of $(t_0,p)$ intervals.}
\label{fig4}
\end{figure}

Figure~\ref{fig3} shows how we select climate data over the same period \emph{$(t_0,p)$} for different years 
and build a corresponding matrix $X(t_{0},p)$ that allows for a SVD analysis: (i) We select 
data from $k$ climate variables over the years, always starting at $t_{0}$ and ending $p$ days later. (ii) 
We stack and normalize the data associated with  year $l$ in a block matrix $\mathcal{B}_l(t_{0},p)$, for $l=1,2,\cdots, N$.
(iii) Finally, all blocks are reshaped into column vectors, forming a new matrix ${\bf X} = {\bf X}(t_0,p)$, which yields
\begin{eqnarray}
\label{eq:schemeP}
	{\bf X}(t_{0},p) = {\bf U} {\bf \Sigma} {\bf V}^{T}  (t_{0},p).
\end{eqnarray}
The columns of ${\bf U}$ -- the SVD modes -- form an orthogonal basis for the space generated by the columns of 
$X$ and the projections of the principal components are given by the ${\bf \Sigma V}^{T}(t_0,p)$ matrix (see Fig~\ref{fig4} \textbf{i}). 

\newpage

\begin{figure}[!h]
\centering
\includegraphics[width=.9\textwidth]
{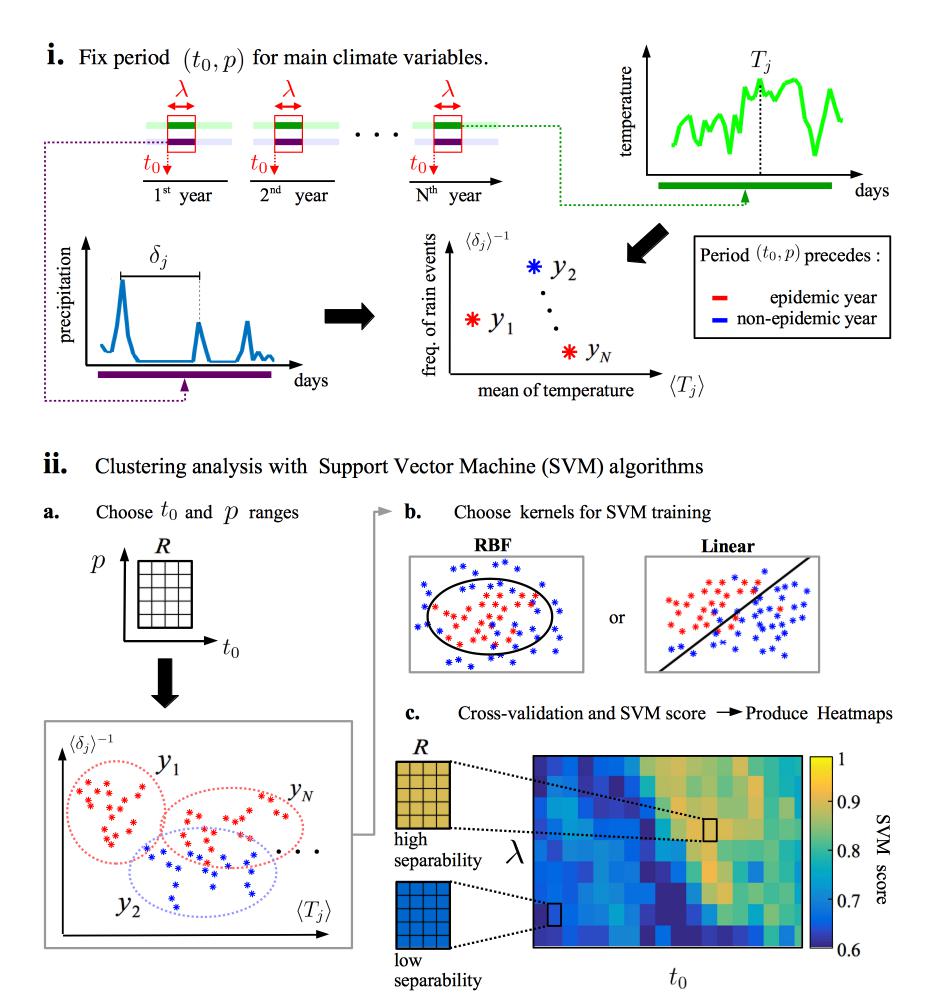}
\caption{{\bf Outline of SVM methodology.} A supervised learning technique for classification:
\textbf{(i)} We calculate and plot mean of average temperature $\left\langle T_j\right\rangle $ and frequency of rain events $\left\langle \delta_j\right\rangle^{-1} $ for a fixed $(t_0,p)$ interval of all years,  using red and blue colors for periods preceding epidemic and  non-epidemic years respectively.
\textbf{(ii)(a)} For each $(t_0,p)$ interval of the rectangle $R$ , we apply (i) to obtain a \emph{cloud} (dashed circles) of points in the plane, for each year. \textbf{(b)} Linear and RBF kernels are used to execute the SVM train/test  and cross validation routines. \textbf{(c)} the SVM score for  $R$ is obtained. We plot $t_0 \times p$ Heatmaps with Regions of High and Low separability scores, which indicates where temperature and precipitation are better correlated with Dengue fever outbreaks.}
\label{fig5}
\end{figure}

In our analysis, we project climate data collected  over \emph{$(t_0,p)$} each year onto a 2-mode plane and label 
years as epidemic (red) or non-epidemic (blue) according to our outbreak convention (see Fig~\ref{fig4} \textbf{ii}). 
This yields a set of $l$ points (one for each year) and allow us to quantify how separate the blue/red dots 
are from each other: we consider two convex hulls connecting red/blue vertices and evaluate the distance  
$\mathcal{H}(t_0,p)$ between them (see Fig~\ref{fig4} \textbf{iii a,b}). Finally, we explore a large range
of values for $t_{0}$ and $p$ to find periods along the yearly cycle in which discrepancies between climate 
conditions might have contributed to dengue outbreaks in the following year.

\subsubsection*{Separability scores from SVM methodology}
\begin{figure}[!t]
\includegraphics[width=.9\textwidth]
{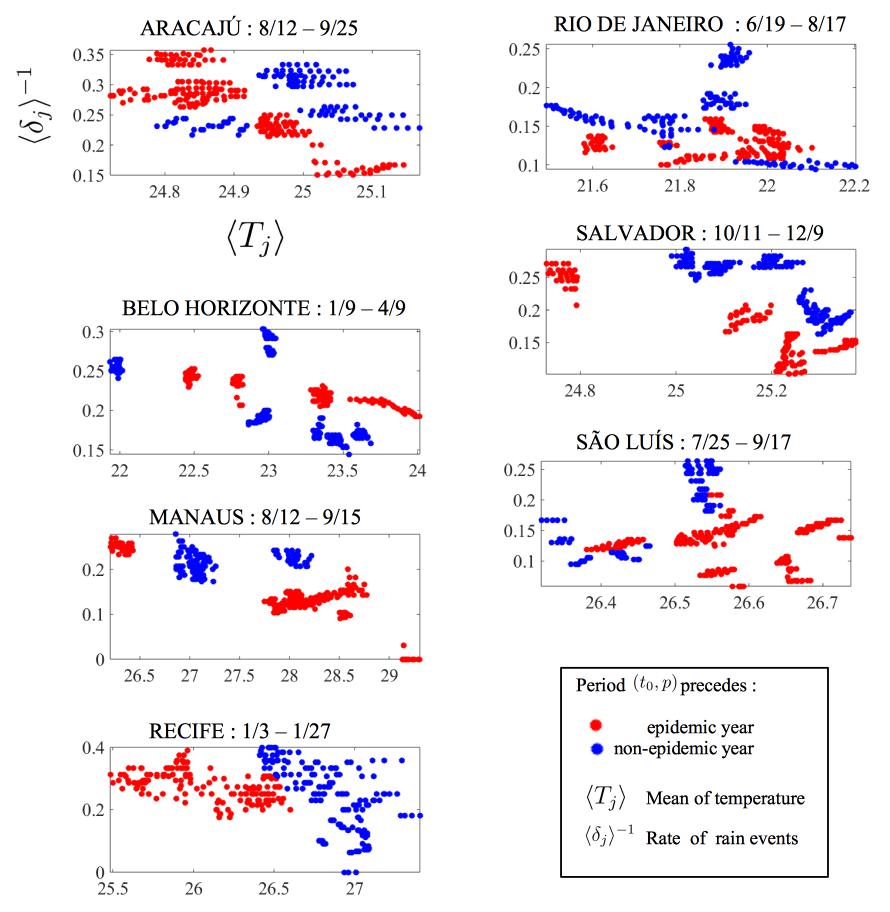}
\caption{{\bf Examples of high separability plots.} For each state Capital we have selected special time windows in which 
there is a clear separation between climate signatures preceding epidemic and non-epidemic years.  Note the distinct 
separation of the data for each individual city, suggesting that a universal model for climate effects across all cities may 
be unattainable.  The separability of data further suggests that epidemics may be accurately predicted in a given capital 
six to nine months in advance of the their outbreak. This separability notion is made quantitatively precise by the SVD and 
SVM separability scores (see text for details). 
}
\label{fig11}
\end{figure}
Our second separability score for measuring discrepancies between climate conditions 
in epidemic/non-epidemic years is based on a supervised learning technique for classification. Figure~\ref{fig5} outlines the main 
steps of our Support Vector Machines (SVM) algorithm: (i) For a fixed $(t_0,p)$ interval, we evaluate two 
climate indicators -- the arithmetic mean of the average temperature $\langle T_j \rangle$  
and average frequency of rain events $\langle \delta_j \rangle ^{-1}$, where 
$\delta_j$ represents time intervals between consecutive peaks on precipitation 
data (see Fig ~\ref{fig5} \textbf{i}). (ii) We label the climate indicators in a 2D plot
as an epidemic year (red) or as a non-epidemic year (blue) according to our
Dengue outbreak criteria. (iii) We repeat the process for $t_0$ and $p$ within a rectangular 
range $R$ in the parameter space. Then, instead of a single point representing year $l$, we have a 
collection of red/blue points (dashed ellipses in fig ~\ref{fig5} \textbf{iia}). In our simulations, the rectangular 
range R was $5 \times 6$, i.e, spanning 5 consecutive starting dates and 6 consecutive duration 
lengths. We tried both a linear kernel and a Radial Basis Function (RBF) kernel for the SVM training 
step on $R$ and cross-validated the climate indicators by sampling 80\% of each dataset 
and testing the accuracy of the predictions in the remaining 20\%. Our separability score 
is ultimately defined as the average classification accuracy after resampling and testing data
for 100 trials.
%
%
\section*{Results}
%
%
\subsection*{Survey of critical climate conditions for different cities} 
In this section, we highlight significant differences between climate conditions 
during epidemic/non-epidemic years for a period starting at day $t_{0}$ and 
duration of $p$ days along the yearly cycle. We postulate
that periods with high separability scores might be of critical importance to the cycle 
of the urban mosquito population and consequently, to the occurrence of dengue 
outbreaks in the following year. The values of $t_{0}$ range from June $1^{st}$
to February $21^{st}$ and the values of $p$ range from $10$--$100$ days, which
completely covers plausible periods that may influence Dengue outbreaks. 
The interpretation of the colormaps presented bellow should be straightforward 
and we highlight (in green) periods/epochs with high separability for both SVD and SVM
methodologies. We restrict our SVD analysis to the five principal modes and fix the color bar for
$\mathcal{H}(t_0,p)$ between 0 and 6 (highest value found for all simulations).  
For the SVM colormaps, we focus our analysis on the highest separability scores and 
choosing  scores above 0.8 for the Linear Kernel. For the RBF kernel, which usually has 
a better predictive performance, the highlight threshold is 0.95.

\begin{figure}[!t]
\includegraphics[width=.9\textwidth]
{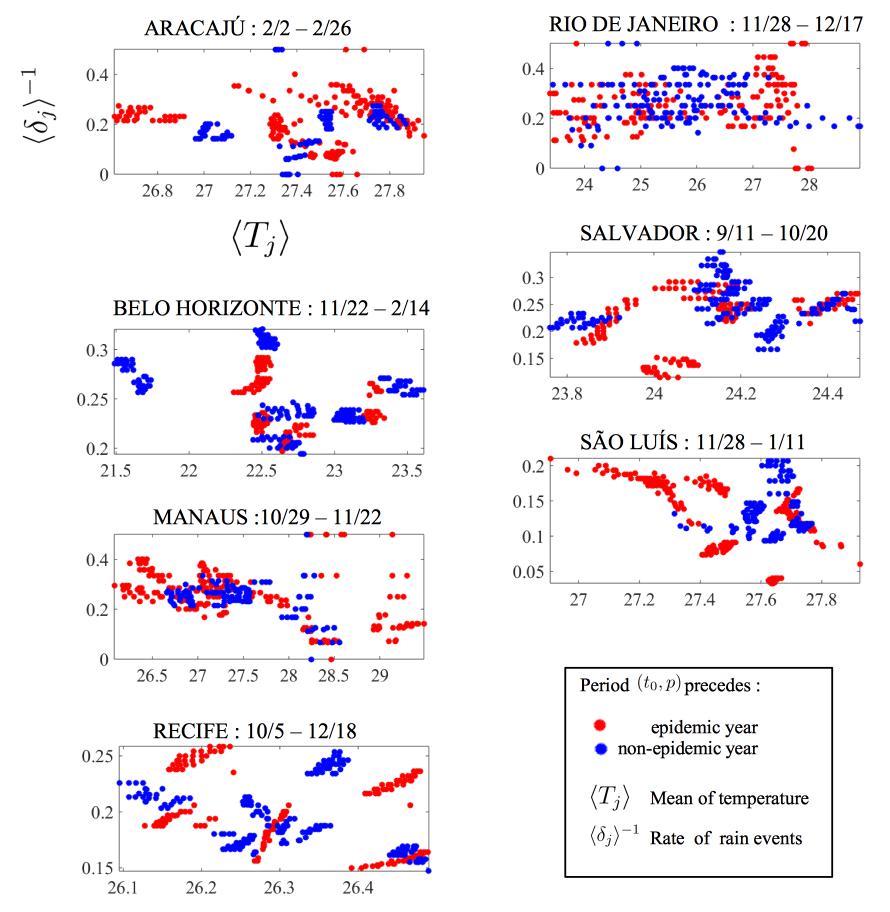}
\caption{{\bf Examples of low separability plots.} Specific time windows in which the epidemic and non-epidemic climate 
variables seems to be poorly distinguishable, therefore not suitable for Dengue prediction.  Unlike Fig.~\ref{fig11}, the mixing 
of data suggests poor predictability across all cities. This separability notion is made quantitatively precise by the SVD and 
SVM separability scores (see text for details). 
}
\label{fig12}
\end{figure}

Figures~\ref{fig4} and \ref{fig5} demonstrate how assessments and scoring are performed for
both the SVD and SVM based methods.   In what follows, a detailed evaluation is made for
each capital city.  Before proceeding to this analysis, however, it is highly informative to interpret
that a high score or low score achieves for separating epidemic and non-epidemic correlations.
Figures~\ref{fig11} and \ref{fig12} demonstrate the clustering of data, or lack thereof,  for all cities. 
In Fig.~\ref{fig11}, representative data for windows achieving a high correlation score is shown.
Remarkably, the red (epidemic) and red (non-epidemic) dots are well separated and distinguishable 
from visual inspection.  Indeed, one could easily postulate decision regions which properly identify, months
in advance, the oncoming presence of a dengue epidemic by simply considering the mean temperature
and precipitation frequency.   Figure~\ref{fig12} shows the data structure when a low correlation
score is achieved.  Note that in this case, there is significant overlap between the red and blue dots, suggesting
that this region for prediction of an epidemic is highly suspect.  Figures~\ref{fig11} and \ref{fig12} provide
an easily interpretable understanding of the predictive nature of our proposed analysis.  It also highlights
important and significant differences between the various Brazilian cities.  Some cities are on the coast, 
while others are in the interior, but regardless, each city has a unique pattern of clustering that can be capitalized
on in order to provide predictive metrics for epidemic outbreaks.    In the figures that follow, a principled 
analysis is performed for each Brazilian city in order to compute regions that give high scores on the SVD/SVM
metrics and provide strong predictive metrics.

\begin{figure}[!t]
\includegraphics[width=1\textwidth]
{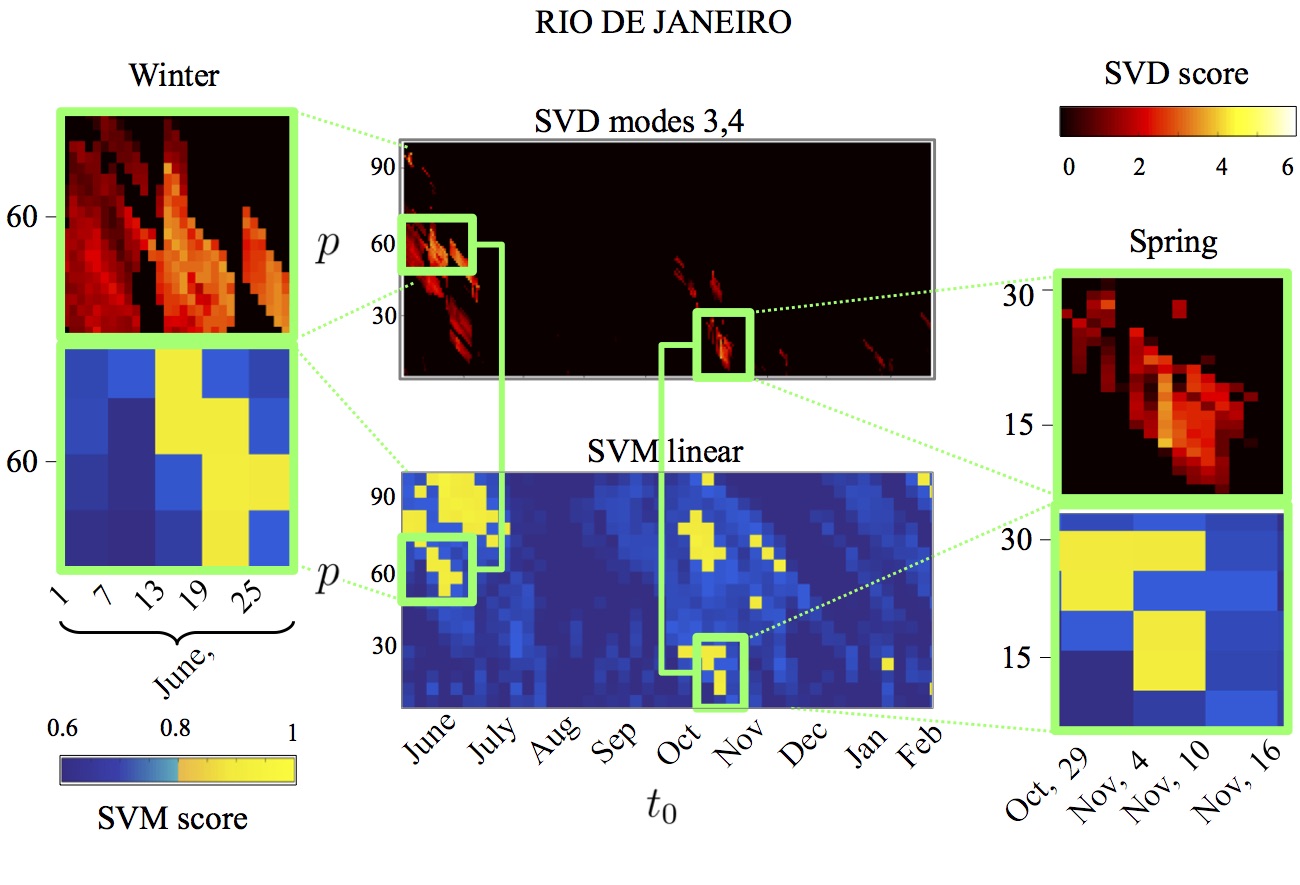}
\caption{{\bf Critical periods for Rio de Janeiro.} There is a good match between the 
different data-driven methods suggesting that specific climate conditions during winter 
season may be crucial to Dengue epidemics. Both methods also indicate a critical period 
of approximately 15 days during spring.}
\label{fig6}
\end{figure}

%
\subsubsection*{Rio de Janeiro}
%
%
Figure ~\ref{fig6} shows periods with high separability scores for the city of Rio de Janeiro. 
Notice that both SVD and SVM methodologies highlight critical epochs during the winter. 
In fact, there is a good accordance between the projection of climate data to $3^{rd}$ 
and $4^{th}$ SVD modes and the linear SVM kernel for $t_0$ in June and $p$ around 60 
days. This suggests that time series for temperature and precipitation from June to August 
may be crucial for the occurrence of Dengue outbreak the following year. There is also good 
accordance between both criteria during the spring, for $t_0$ starting in October-November 
and $p$ around 15 days.

\begin{figure}[!t]
\includegraphics[width=1\textwidth]
{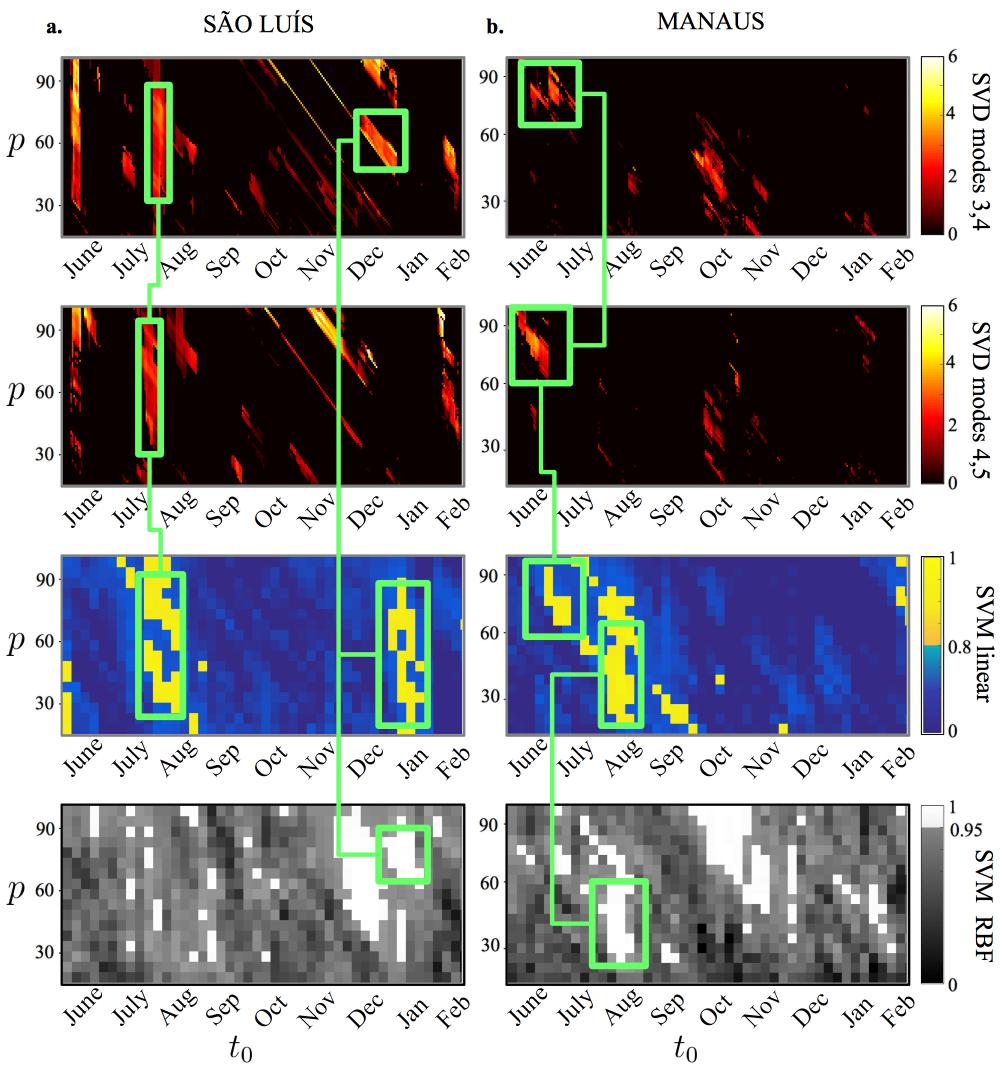}
\caption{{\bf Critical periods for S\~ao Lu\'is and Manaus.} The two capitals in the north of Brazil exhibit 
good accordance between SVD separability scores (for modes 3,4 and 4,5) and SVM separability scores
(for both Linear and RBF kernels). \textbf{a.} Temperature and precipitation are correlated with Dengue 
outbreaks during winter and summer in the case of S\~ao Lu\'is. \textbf{b.} For  Manaus, the correlation 
is higher during winter and spring.}
\label{fig7}
\end{figure}

\begin{figure}[!t]
\includegraphics[width=1\textwidth]
{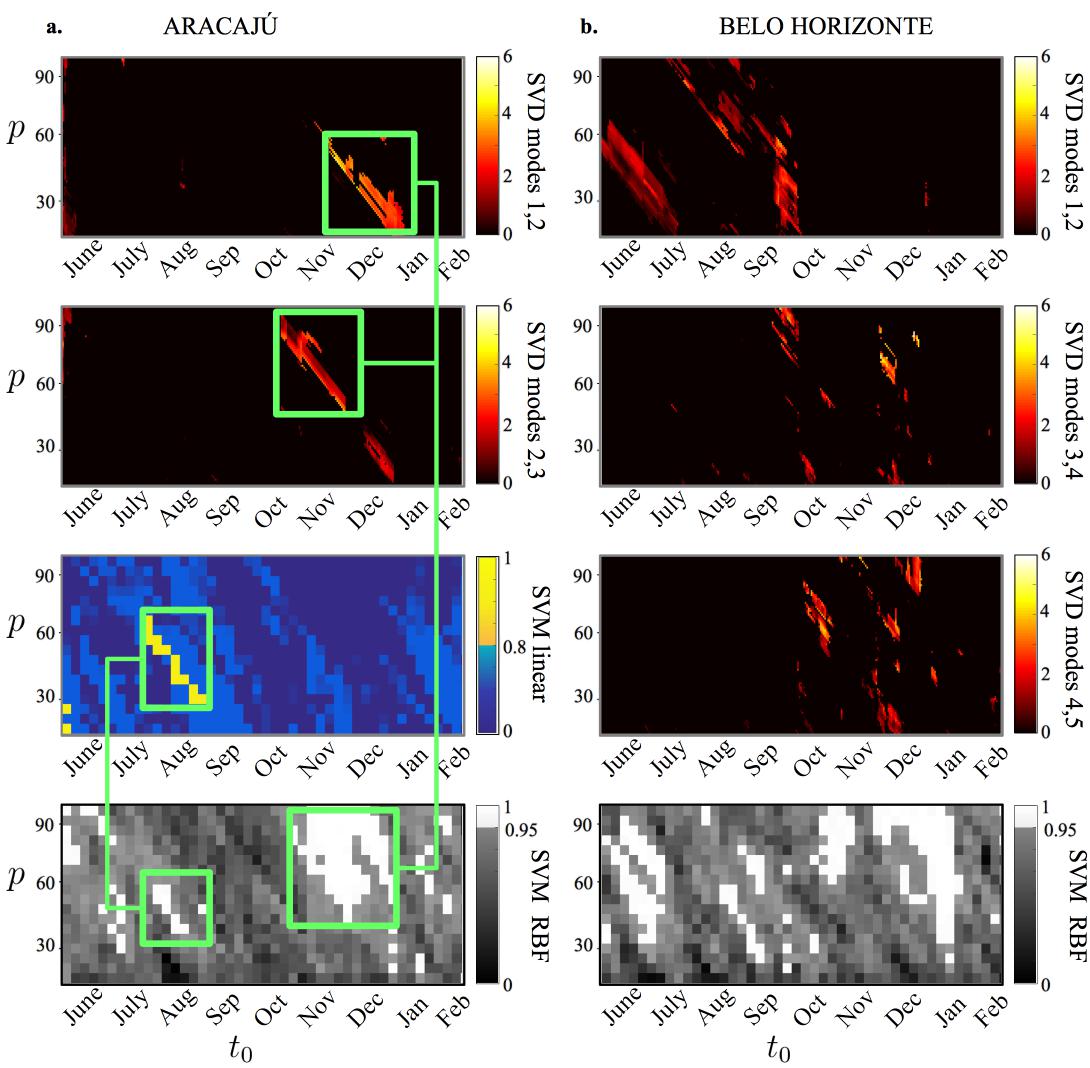}
\caption{{\bf Critical periods for Aracaj\'u  and Belo Horizonte.} For these cities, we have found  periods with 
high correlation between climate indicators and Dengue outbreaks during winter, spring and summer. 
Aracaj\'u (\textbf{a.}) and Belo Horizonte (\textbf{b.}) are the state capitals of Sergipe and Minas Gerais, 
located in the northeast and southeast regions of Brazil, respectively.}
\label{fig8}
\end{figure}
%

%
%
\subsubsection*{S\~ao Lu\'is}
%
%
Figure ~\ref{fig7}\textbf{a.} shows critical periods for S\~ao Lu\'is, the state capital of Maranh\~ao.
Overall, we found good accordance between separability scores provided by the SVM and 
SVD methods. The  SVD method indicated (for modes 3,4 and 4,5) critical $(t_0,p)$ intervals 
for $t_0$ in July and $p$ varying from 30 to 85 days. A similar period was found with the SVM 
method (using a linear kernel). This match suggests that temperature and rain in late winter 
and beginning of spring may play an important role in the occurrence of Dengue outbreaks. 
Another critical period indicated by both methods has $t_0$ in December and duration $p$ 
around 60 days. 
%
%
\subsubsection*{Manaus}
%
%
\noindent The capital of Amazonas has a set of periods with 
high separability scores 
in the winter (see Fig ~\ref{fig7}\textbf{b}). In the SVD colormaps (for modes 3,4 and 4,5), the 
separability score is high for $t_0$ between June and July and $p$ between 60 and 90 days. 
This corresponds to the months of June, July and August. This is in good accordance with 
the scores given by the SVM methodology (using a linear kernel). We also highlight that 
there is a good match between the methods during a critical period with $t_0$ lying between 
July and August and $p \leq 60$ days, which would also include the first days of spring.
%
%
%
%
%
\subsubsection*{Aracaj\'u}
%
%
The capital of Sergipe displays high separability scores according to the SVD methodology
(for modes 1,2 and 2,3) for periods with $t_0$ in November -- January and period length 
$p<60$ days (see Fig ~\ref{fig8} \textbf{a}). Similar critical periods in the $t_0 \times  p$-plane 
also occur in the SVM-RBF color map. This suggests that the climate conditions during spring 
and summer are of crucial importance for the occurrence of Dengue in Aracaj\'u. The SVM 
methodology also highlights critical periods for $t_0$ in August and $p$ between 30 and 60 days, 
which would correspond to late winter and/or beginning of spring. 

\begin{figure}[!t]
\includegraphics[width=1\textwidth]
{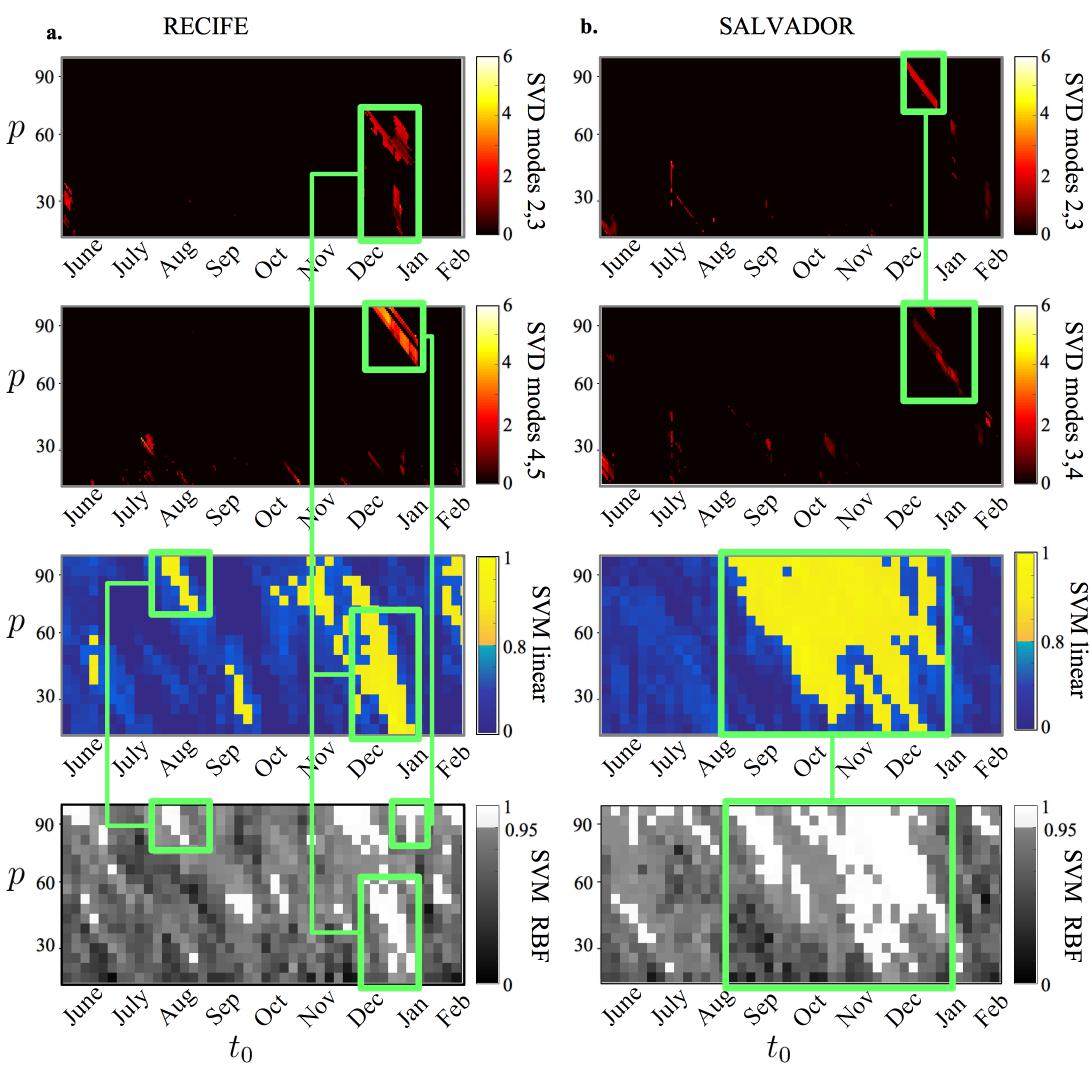}
\caption{{\bf Critical periods for Recife and Salvador.} These two northeast capitals have exhibited strong correlation between climate signatures and Dengue epidemics, specially during spring and summer. \textbf{a.} For Aracaj\'u, we have found accordance between SVD (modes 2,3 and 4,5) and the SVM methods. \textbf{b.} For Salvador, SVM methods has shown a good performance by showing big RHS for $t_0$ between August and December.}
\label{fig9}
\end{figure}

\subsubsection*{Belo Horizonte }
%
%
In Fig ~\ref{fig8} \textbf{b.} we show the highlights for Belo Horizonte, the state capital of 
Minas Gerais. The SVD color map (for modes 1,2) shows critical regions for $t_0$ between 
June--July and $p$ between 30 and 60 days. These $(t_0,p)$-periods corresponds to the winter season in Brazil. A similar result was found in the SVM - RBF method, but with a larger range of $p$. There was also a good accordance between SVD scores (for modes 4,5) and SVM (for the RBF kernel) when 
$t_0$ is between October and November and $p$ is between 60 and 90 days. These 
critical periods with high separability scores correspond to the spring season. For the summer period and beginning of the fall (where the epidemic outbreaks usually occur), both SVM kernels indicate critical 
periods for $t_0$ between December and January and $p$ between 45 and 90 days. 
\subsubsection*{Recife}
The capital of Pernambuco shows high separability scores for both 
SVD and SVM methodologies during the summer season (see Fig ~\ref{fig9}\textbf{a}).
We found critical periods for $t_0$ between December--January and $p$ varying from 
15 to 60 days using the SVD methodology (for modes 2,3) and the SVM methodology
(for both Linear and RBF kernels). Both SVD color map (for modes 4,5) and SVM 
color map (RBF kernel) indicate regions with high separability scores for 
$t_0$ between December and January and $p$ around between 60 and 90 day, which
would include the first days of the fall season. For winter and spring, the SVM methods
(for both linear and  RBF kernels) find critical periods for $t_0$ in August and $p$ 
around 90 days.
\subsubsection*{Salvador}
For the state capital of Bahia (see Fig ~\ref{fig9} \textbf{b}), the SVM separability 
scores are high for $t_0$ between August and December (for both  Linear and RBF kernels) 
and  for all values of length $p$.  This suggests that spring and summer are crucial for 
the development of Dengue epidemics in Salvador. We also highlighted that SVD separability
scores (for modes 2,3 and 3,4) are high for $t_0$ between December and January and
$p$ between 60 and 90 days, which would also correspond to the summer season.
%
%
%

%
%
\section*{Discussion}
In this work, we developed data-driven methods to identify in a systematic manner a set of 
critical periods in the annual cycle in which climate conditions may play a significant role 
in the development of Dengue outbreaks the following year. For a fixed time period starting at $t_{0}$ 
and lasting $p$ days, we evaluate separability scores between the climate conditions on 
epidemic/non-epidemic years. We postulate that the periods where these climate conditions differ 
most might be crucial for the development of the life cycle of the mosquito population, and consequently, 
to Dengue outbreaks. 
The separability scores were calculated following two different methods. The first one 
is based on dimensionality reduction of data via Singular Value Decomposition (SVD) and the second one 
on the machine learning classification algorithm known as Support Vector Machines (SVM).   
We applied these methods to temperature and precipitation time series data for seven state capitals in 
Brazil where there was a significant alternation between epidemic and non-epidemic years in 
the recent past. Both methods indicated critical periods with remarkable agreement. 
The analysis of this particular dataset was only made possible due to the successful application 
of compressed sensing techniques to plausibly complete missing data. In fact, the cities of 
Rio de Janeiro, Salvador, and São Luís had the larger gaps in their daily recording of climate
variables that were circumvented using compressive sensing.

After localizing the critical periods with high separability scores between epidemic/non-epidemic climate conditions
we were able to find which seasons were crucial for the development of Dengue outbreaks at each city. See 
Table ~\ref{table1} for a summary of the results. We obtained strong evidence that the climate influence on epidemics 
varies significantly from place to place ~\cite{Adde,Liao2014,Johansson2009}, and thus rejecting simplistic or universal 
explanations involving temperature and rain precipitation in urban centers. We found a high correlation between critical 
climate signatures during the winter season and the occurrence of outbreaks in Aracajú, Belo Horizonte, Manaus, 
Rio de Janeiro, and S\~ao Luís. 

Several works report and quantify how climate influence the mosquito development on a weekly scale 
~\cite{Pessanha,Honorio2009,Dibo}. We suggest that climate conditions may have long-term effects as well, 
occurring even months before the outbreaks. As a consequence, intensifying mosquito control campaigns 
during the winter season may prove an interesting epidemic control strategy, especially due to the smaller size
of the vector populations during that period. In Brazil, the national and local campaigns are usually restricted to spring 
and summer periods~\cite{dengue2014,dengue2016}. In fact, the Brazilian government announced that a 
special task force for fighting mosquitos was to be formed November $3^{rd}$, 2016~\cite{dengue2017}. We believe
this starting date to be too late since critical climate conditions were detected in some cities even 9 months prior 
to epochs with higher Dengue incidence. 
\begin{table}[!t]
\centering
\caption{
{\bf Summary of most important seasons  for dengue outbreaks.}}
\begin{tabular}{|c|c|c|c|c|c|}
\hline
\multicolumn{1}{|l|}{\bf Capital} & \multicolumn{1}{|l|}{\bf Winter } & \multicolumn{1}{|l|}{\bf Spring }&\multicolumn{1}{|l|}{\bf Summer }&\multicolumn{1}{|l|}{\bf Fall (DF)  }\\ \hline 
 Aracajú &  x &  &x&      \\ \hline
Belo Horizonte  &  x&x  &x&       \\ \hline
Manaus  &x   &  & &     \\ \hline
Recife  &   &x  &x  &x      \\ \hline
Rio de Janeiro  & x  & x & &      \\ \hline
Salvador  &   &x  &x &      \\ \hline
São Luís  &x   &  &x &      \\ \hline
\end{tabular}
\begin{flushleft} 
\small{Remark: Peaks of Dengue Fever outbreaks happen typically during the fall (March -- May).}
\end{flushleft}
\label{table1}
\end{table}
\begin{figure}[!t]
\includegraphics[width=1\textwidth]
{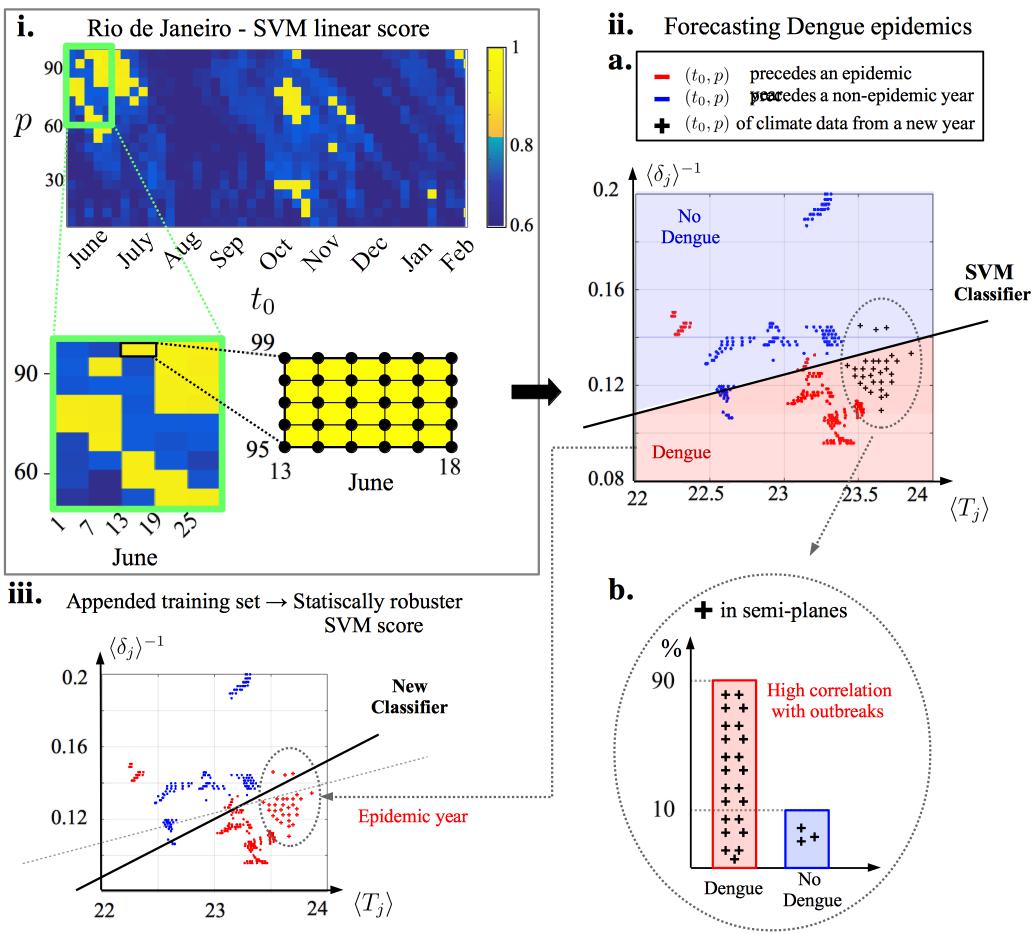}
\caption{{\bf Forecasting Dengue Outbreaks and appending data for further analysis.} Example for  the SVM-Linear methodology on climate data from Rio de Janeiro. \textbf{(i)} We choose a high scored $(t_0,p)$-rectangle, for which we plot the climate indicators  with their respective colors. \textbf{(ii)} We apply a SVM training algorithm on this 2D-dataset. \textbf{(a)}  A classifier line can be drawn and two semi-planes (Dengue and No-Dengue) are obtained. \textbf{(b)}  With data from a new year for the same $(t_0,p)$ periods (black crosses), we can compute the percentage of indicators that falls into each of those semi-planes. Therefore we are able to estimate the correlation between new and previous climate data with respect to Dengue epidemics. \textbf{(iii)} Depending on the classification of the new year as epidemic or not,  the new data is colored red or blue to become part of a new SVM-training set. This procedure will give a more accurate information about the importance of the chosen $(t_0,p)$-rectangle on Dengue prediction.}
\label{fig10}
\end{figure}

A number of early warning systems are available for calculating the risk of Dengue epidemics taking climate factors into 
account~\cite{Buczak2012,Buczak2014,Lowe2011,Lowe2012, Racloz2012, Johansson2016}. In this sense, our methodology offers an additional set of key periods
that may assist current warning systems or serve as basis to a new model focusing on climate signatures of epidemic years.  
Figure ~\ref{fig10} illustrates how this could be achieved using SVM linear kernel separability scores: (i) We would train climate 
data projected onto the $\langle T_j \rangle  \times  \langle \delta_j \rangle ^{-1}$ plane and divide it into two regions referring
to epidemic (red) or non-epidemic (blue) data. (ii) Once we obtain temperature and precipitation measurements for the following
year we can quantify the fraction of climate data falling into each region and use it to forecast Dengue the following year. 
(iii) After the Dengue outcome is known for that year, we can append that data to our set and retrain the classifier line between 
epidemic/non-epidemic regions. This should improve, at least in theory, the precision of future forecasts and our understanding 
of critical climate signatures.

There are several limitations to our work and all of our results must be interpreted with caution 
and parsimony. Ultimately, we are only \textit{suggesting} that temperature and precipitation discrepancies in key epochs affected 
mosquito populations in a critical way, and despite the plausible hypotheses, they were not yet directly measured or reported satisfactory by field studies. Moreover, we didn't consider several other factors believed to be important for explaining 
dengue dynamics in details, such as: \textbf{(i)} Circulation of different strains of the dengue virus ~\cite{Rabaa,Raghwani,Diaz,Adams}; once the cross-immunity wanes with time, the introduction of new DENV serotypes may affect an entire population.  \textbf{(ii)} Human mobility within and among the cities ~\cite{manbites,stoddard,stolerman,Wesolowski,Barmak}; the lifetime movement range of an Ae.Aegypti mosquito is 
typically less than a kilometer and the spread of Dengue through an urban area is most likely driven by everyday human movement
~\cite{Harrington,Medeiros}. In fact, humans act as vectors between relatively localized mosquito populations 
and might change their commute strategies based on climate factors as well. \textbf{(iii)} Human demographic dynamics:
lower death rates may increase the longevity of immune individuals and lower birth rates may decrease the number of susceptible individuals. Such fluctuations over the years may change the magnitude of the infections ~\cite{Cummings,Mondini}.
\textbf{(iv)} Global warming and global climate changes: several studies examined for instance the influence of El Niño Southern Oscillation (ENSO) in dengue incidence. While some argue that ENSO is behind the synchronization of dengue epidemics and traveling waves of infections, others dismiss it as a minor factor~\cite{MAJohansson, Cazelles, Banu,Naish }. In any case, global climate changes are likely to affect local climate conditions and consequently, 
dengue transmission.  
\textbf{(v)} Our methodological limitations constrained the analysis to cities that experienced at least three epidemic and three non-epidemic years in the recent past. Poorly recordings of climate data also prevented us from including two additional capitals to our dataset
(see Appendix and SI for details). In the future, we expect to extend our analysis to other capitals. Finally, at this stage, we outline a modest 
prediction system for Dengue outbreaks using our methods only as a proof of concept, leaving detailed forecasting (as done in \cite{Buczak2012,Buczak2014}) for future works.

\section*{Conclusion}
Epidemic control of Dengue, Zika and Chikungunya is one of the most urgent public health 
challenges in a globalized world, and their effects have dramatic societal consequences in large 
tropical countries such as Brazil.  A better understanding of the the multi-scale and long terms effects 
of climate conditions on the development of {\em Aedes Aegypti} populations is crucial for improving the timing 
of vector-control efforts and other policies. In this sense, this work adds a new piece to the complex puzzle 
that is the development of mosquito populations in dynamic urban environments under variable climate conditions. Figures~\ref{fig11} and~\ref{fig12} illustrates the power of our methodology. We have found two specific parameters -- mean of temperature and frequency of precipitation -- that may be crucial for Dengue prediction in Brazil.  

Not only are the analytic metrics developed in this manuscript predictive, they are also easily interpretable in terms of simple to acquire measurement proxies of temperature and precipitation.  Moreover, in all cities considered where data was readily available, distinct and separable climate patterns were shown which suggest that accurate
prediction of epidemics can be achieved in the winter preceding the outbreak.  This suggests that
many of the eradication strategies should be performed well in advance of the summer months where
the epidemic is manifest.  

Table~\ref{table1} summarizes the potential for predictive success of 
dengue outbreaks.  Remarkably, in almost all cities, aside from Recife
and S\~ao Lu\'is, a prediction can be made approximately six to nine months in advance of the epidemic outbreak.  And aside from Manaus, all cities offer multiple windows of opportunity for forecasting the dengue levels during the annual cycle.  Interestingly, the summer in Rio de Janeiro offers little insight into this matter, since data of years with and without dengue are qualitatively similar from a climate perspective.     Yet public strategies have typically been enacted and decided during this time period, which is both too late and does not leverage the predictive capabilities of the climate data.   We conjecture that the winter months are critical for establishing the ideal breeding conditions, through temperature and frequency of precipitation, which ultimately determine the size of the {\em Aedes Aegypti} population.  This suggests that disrupting the breeding cycle six to nine months in advance may be a robust strategy for vector control.  For instance, in Rio de Janeiro, if during the winter months the rain frequency is approximately once per week and the average temperature is approximately 22 Celcius (See Fig.~\ref{fig11}), then it is highly likely that an epidemic will occur, thus requiring an intervention strategy 9 months in advance.

The work also highlights that the patterns allowing for predictive success are quite distinct from
city to city.   Figure~\ref{fig11} demonstrates that a simple, universal rule about climate effects may be hard
to achieve.  Indeed, data on the seven cities demonstrate a remarkably heterogeneous range of behaviors despite
each individual city giving rise to clear prediction windows.  This is largely to be expected as climatic
effects, such as proximity to ocean, jungle, forest, dense populations, etc. will all play a significant
role in how precipitation and temperature favorably or unfavorably effects the growth of the disease
vector {\em Aedes Aegypti}.

\section*{Acknowledgments}
LS, PDM and JNK would like to acknowledge Prof. Stefanella Boatto from 
Federal University of Rio de Janeiro (UFRJ), Joshua L. Proctor from the Institute of Disease Modeling, and Prof. Roberto.I. Oliveira (IMPA) for their insightful comments and 
enthusiastic support of this work. A significant part of this project was done during
Lucas' internship at the University of Washington in the winter of 2016 funded by 
the Brazilian National Council of Research (CNPq).

\newpage
\section{Supporting Information}

\paragraph*{S1 Appendix}
\label{S1_Appendix}\textbf{Details about the choice of the seven capitals}
As explained in our Methods section, we chose capitals that had at least 3 years with Dengue Epidemics (DE) and at least 3 years without DE in the recent past. 
The following 9 capitals passed this criterium: Aracaj\'u, Belo Horizonte, Cuiab\'a, Jo\~ao Pessoa, Manaus, Recife, Rio de Janeiro, Salvador and S\~ao Lu\'is. We completed missing data through linear interpolation and/or usage of alternative sources for precipitation time series given that the CVX routine does not work well for episodical data events. From the 9 capitals, the following 6 had only single precipitation gaps: Aracaj\'u, Belo Horizonte, Manaus, Recife, Salvador and S\~ao Lu\'is. The cities of Cuiab\'a, Jo\~ao Pessoa and Rio de Janeiro had big missing data epochs. For Rio de Janeiro we found an alternative source of precipitation data, but the other two capitals had to be discarded from our analysis. 
\vspace{1em}

\paragraph*{S1 Tables}
\label{S1_Tables}
{\bf Epidemic and non-epidemic years for each chosen capital and missing Climate data for each chosen capital}
We provide tables with estimated population, total number of Dengue cases, incidence per $100,000$ inhabitants, and details of our 
climate data completing protocols (if any). For Rio de Janeiro, we consider the time period from 2003 to 2013 and we use epidemic data from municipal 
webpage (\href{http://www.rio.rj.gov.br/web/sms/dengue}{link}). For the other capitals, the analyzed period ranges from 2002 to 2012 and data 
was collected from the Ministry of Health's  Notifiable Diseases Information System. (\href{http://www.portalsinan.saude.gov.br}{SINAN})

\begin{table}[!ht]
\centering
\caption{
{\bf Aracaj\'u.}}
\begin{tabular}{|l|l|l|l|l|l|l|}
\hline
\multicolumn{1}{|l|}{\bf Year} & \multicolumn{1}{|l|}{\bf Pop.} & \multicolumn{1}{|l|}{\bf Cases}& \multicolumn{1}{|l|}{\bf Incidence}\\ \hline 
$2002$ & $473,991$ & $1,933$  & \cellcolor{red!25}407.81   \\ \hline
$2003$ & $479,767$ & $1,301$  &  \cellcolor{red!25}271.17 \\ \hline
$2004$ & $491,898$ & $ 166$   &  \cellcolor{blue!25} 33.75 \\ \hline
$2005$ & $498,619$ & $ 271$   &  \cellcolor{blue!25} 54.35  \\ \hline
$2006$ & $505,286$ & $ 355$   &  \cellcolor{blue!25}70.26\\ \hline
$2007$ & $520,303$ & $728$    & \cellcolor{red!25}139.92\\ \hline
$2008$ & $536,785$ & $1,0702$ & \cellcolor{red!25}1,993.72\\ \hline
$2009$ & $544,039$ & $1,232$  & \cellcolor{red!25}226.45\\ \hline
$2010$ & $571,149$ & $302$    & \cellcolor{blue!25}52.88\\ \hline
$2011$ & $579,563$ & $1,399$  & \cellcolor{red!25}241.39\\ \hline
$2012$ & $587,701$ & $2,656$  & \cellcolor{red!25}451.93\\ \hline
\end{tabular}
\begin{flushleft} 
\vspace{.5cm}
Incidence = Cases per $100,000$ inhabitants. Single gaps of missing climate data were filled by linear interpolation; temperature on 12/21/2006 and 
precipitation on 7/24/2006.  
\end{flushleft}
\label{table3}
\end{table}

\begin{table}[!ht]
\centering
\caption{
{\bf Belo Horizonte.}}
\begin{tabular}{|l|l|l|l|l|l|l|}
\hline
\multicolumn{1}{|l|}{\bf Year} & \multicolumn{1}{|l|}{\bf Pop.} & \multicolumn{1}{|l|}{\bf Cases}&\multicolumn{1}{|l|}{\bf Incidence}&\multicolumn{1}{|l|}{\bf Temp (L.I)}&\multicolumn{1}{|l|}{\bf Precip (L.I)} \\ \hline 
$2001$ &    --         &     --         &     --                & 8/9   & 8/9  \\ \hline
$2002$ & $2,284,468$ & $4,749$  &   \cellcolor{red!25}207.88 & 8/31   & 8/31  \\ \hline
$2003$ & $2,305,812$ & $ 1,800$  &  \cellcolor{blue!25}78.06 &  --        & -- \\ \hline
$2004$ & $2,350,564$ & $472 $   &  \cellcolor{blue!25} 20.08 &  --      & -- \\ \hline
$2005$ & $2,375,329$ & $149 $   &  \cellcolor{blue!25} 6.27  &   --       & --\\ \hline
$2006$ & $  2,399,920$ & $872 $   & \cellcolor{blue!25} 36.33&   --       & --\\ \hline
$2007$ & $  2,412,937$ & $ 5278$    &  \cellcolor{red!25} 218.74 & 12/31   & 12/31 \\ \hline
$2008$ & $ 2,434,642$ & $12,967$ & \cellcolor{red!25}532.60  &  1/1    &  1/1  \\ 
       &              &          &    \cellcolor{red!25}     &  11/21  &  11/21  \\ \hline
$2009$ & $   2,452,617$ & $14,494$  & \cellcolor{red!25}590.96 &  12/12    & 12/12\\ \hline
$2010$ & $ 2,375,151$ & $52,315$    &  \cellcolor{red!25} 2,202.60 &  --    & --\\ \hline
$2011$ & $2,385,640$ & $1,749$  &  \cellcolor{blue!25} 73.31 &  --        & --\\ \hline
$2012$ & $2,395,785$ & $635$  & \cellcolor{blue!25}26.50 & --       & -- \\ \hline
\end{tabular}
\begin{flushleft}
\vspace{.5cm}
Incidence = Cases per $100,000$ inhabitants.  L.I stands for Linear Interpolation.
\end{flushleft}
\label{table4}
\end{table}

\begin{table}[!ht]
\centering
\caption{
{\bf Manaus.}}
\begin{tabular}{|l|l|l|l|l|l|l|}
\hline
\multicolumn{1}{|l|}{\bf Year} & \multicolumn{1}{|l|}{\bf Pop.} & \multicolumn{1}{|l|}{\bf Cases}&\multicolumn{1}{|l|}{\bf Incidence} \\ \hline 
$2002$ & $ 1,488,805$   & $  1,855$  &  \cellcolor{red!25}124.60    \\ \hline
$2003$ & $ 1,527,314$   & $3,731$    & \cellcolor{red!25} 244.29  \\ \hline
$2004$ & $ 1,592,555$   & $789 $     &  \cellcolor{blue!25} 49.54 \\ \hline
$2005$ & $ 1,644,690$   & $ 915 $    &  \cellcolor{blue!25}55.63   \\ \hline
$2006$ & $ 1,688,524$   & $495 $   &  \cellcolor{blue!25}29.32 \\ \hline
$2007$ & $ 1,646,602$   & $1,989$    &  \cellcolor{red!25}120.79  \\ \hline
$2008$ & $1,709,010$    & $5,975$ &  \cellcolor{red!25}349.62 \\ \hline
$2009$ & $1,738,641$    & $623$  &  \cellcolor{blue!25}35.83  \\ \hline
$2010$ & $ 1,802,014$   & $3,748$    &  \cellcolor{red!25} 207.99  \\ \hline
$2011$ & $ 1,832,424$   & $54,342$  &  \cellcolor{red!25}2,965.58 \\ \hline
$2012$ & $1,861,838$    & $3,703$  &  \cellcolor{red!25}198.89 \\ \hline
\end{tabular}
\begin{flushleft}
\vspace{.5cm}
Incidence = Cases per $100,000$ inhabit.  Single gaps of missing climate data were filled by linear interpolation; 
temperature on 12/23/2005 and precipitation on 2/11/2005.  
\end{flushleft}
\label{table5}
\end{table}

\begin{table}[!ht]
\centering
\caption{
{\bf Recife.}}
\begin{tabular}{|l|l|l|l|l|l|l|}
\hline
\multicolumn{1}{|l|}{\bf Year} & \multicolumn{1}{|l|}{\bf Pop.} & \multicolumn{1}{|l|}{\bf Cases}&\multicolumn{1}{|l|}{\bf Incidence}&\multicolumn{1}{|l|}{\bf Temp (L.I)}&\multicolumn{1}{|l|}{\bf Precip(L.I)} \\ \hline 
$2001$ &     --         &    --     &             --              &  --   &  --  \\ \hline
$2002$ & $1,449,135 $   & $42,791$  & \cellcolor{red!25}2,952.86  &  --   &  --  \\ \hline
$2003$ & $ 1,461,320 $  & $449$     &  \cellcolor{blue!25}30.73   &  --   &  --  \\ \hline
$2004$ & $  1,486,869 $ & $ 241 $   &  \cellcolor{blue!25}16.21   &  --   &  --  \\ \hline
$2005$ & $1,501,008$    & $ 830 $   &  \cellcolor{blue!25}55.30   &  --   &  --  \\ \hline
$2006$ & $  1,515,052$  & $1,443 $  &  \cellcolor{blue!25}95.24   & 11/4  &  --  \\ 
       &                &           &  \cellcolor{blue!25}        & 12/2  &      \\ \hline
$2007$ & $1,533,580 $   & $ 1,503$  &  \cellcolor{blue!25}98.01   &  --   &  --  \\ \hline
$2008$ & $1,549,980$    & $4,771$   &  \cellcolor{red!25}307.81   & 4/28  &  --  \\ \hline
$2009$ & $1,561,659$    & $578$     &  \cellcolor{blue!25}37.01   & 4/30  &      \\ 
       &                &           &  \cellcolor{blue!25}        & 7/31  &  --  \\ 
       &                &           &  \cellcolor{blue!25}        & 11/19 &      \\ \hline
$2010$ & $1,537,704 $   & $11,494$  &  \cellcolor{red!25}747.48   & 9/8   &  --  \\ \hline
$2011$ & $1,546,516 $   & $5,471$   & \cellcolor{red!25}353.76    &  --   &  --  \\ \hline
$2012$ & $1,555,039$    & $11,444$  & \cellcolor{red!25}735.93    &  1/1  &  1/1 \\ 
       &                &           & \cellcolor{red!25}          &  5/2  &  5/2 \\ 
       &                &           &\cellcolor{red!25}           &  6/14 &  8/14\\ 
       &                &           & \cellcolor{red!25}          &  8/14 &      \\ \hline
\end{tabular}
\begin{flushleft} 
\vspace{.5cm}
Incidence = Cases per $100,000$ inhabitants.  L.I stands for Linear Interpolation.
\end{flushleft}
\label{table6}
\end{table}

\begin{table}[!ht]
\centering
\caption{
{\bf Rio de Janeiro.}}
\begin{tabular}{|l|l|l|l|l|l|l|}
\hline
\multicolumn{1}{|l|}{\bf Year} & \multicolumn{1}{|l|}{\bf Pop.} & \multicolumn{1}{|l|}{\bf Cases}&\multicolumn{1}{|l|}{\bf Incidence}&\multicolumn{1}{|l|}{\bf Temp (CVX)}&\multicolumn{1}{|l|}{\bf Precip (subst)} \\ \hline 
$2002$ & --  & --  & -- & 8/31 & --   \\ \hline
$2003$ & $5,974,081 $ & $1,610$  &  \cellcolor{blue!25} 26.95 &   3/1 -- 3/2   & 6/20 -- 6/30  \\ 
       &              &          &  \cellcolor{blue!25}       &  6/20 -- 6/30  &                \\ \hline  
$2004$ & $6,051,399 $ & $607$  &  \cellcolor{blue!25}10.03 & -- & -- \\ \hline
$2005$ & $ 6,094,183 $ & $980 $   & \cellcolor{blue!25}16.08 & -- & -- \\ \hline
$2006$ & $6,136,652$ & $14,435 $   &  \cellcolor{red!25} 235.23 & -- & 12/13 -- 12/31 \\ \hline
$2007$ & $6,093,472$ & $26,507 $   &  \cellcolor{red!25}  435.01  & 1/1 -- 2/1 & 1/1 -- 1/10   \\ \hline
$2008$ & $6,161,047 $ & $ 110,861$    &  \cellcolor{red!25} 1799.39 & -- & -- \\ \hline
$2009$ & $6,186,710$ & $ 2,961$ &  \cellcolor{blue!25}47.86 &  2/11 & -- \\ \hline
$2010$ & $6,320,446$ & $3,000$  & \cellcolor{blue!25} 47.47 & -- & --\\ \hline
$2011$ & $6,355,949 $ & $78,645$    & \cellcolor{red!25}1237.34 & --  & --\\ \hline
$2012$ & $6,390,290 $ & $137,505$  & \cellcolor{red!25}2151.78  &  12/8             &  --  \\ 
       &              &          &  \cellcolor{red!25}          &  12/26 -- 12/27   &                \\ \hline   
$2013$ & $6,429,923$ & $ 66,278 $  & \cellcolor{red!25}1030.77 &  6/13 -- 6/21 & 6/14 -- 6/19 \\ \hline
\end{tabular}
\begin{flushleft}
\vspace{.5cm}
Incidence = Cases per $100,000$ inhabitants. Number of Dengue cases in 2013 taken from the City's hall health department, 
because data from SINAN is not available for that year. For the larger gaps of missing data on precipitation time series, we have used 
data of the \emph{Alerta Rio} system  from  Sa\'ude neighborhood, the closest to the Santos Dumont airport where INMET's rain 
collectors are located. 
\end{flushleft}
\label{table2}
\end{table}

\begin{table}[t]
\centering
\caption{
{\bf Salvador.}}
\begin{tabular}{|l|l|l|l|l|l|l|}
\hline
\multicolumn{1}{|l|}{\bf Year} & \multicolumn{1}{|l|}{\bf Pop.} & \multicolumn{1}{|l|}{\bf Cases}&\multicolumn{1}{|l|}{\bf Incidence}&\multicolumn{1}{|l|}{\bf Temp (CVX)}&\multicolumn{1}{|l|}{\bf Precip (L.I)}\\ \hline 
$2001$ & --  &  -- & -- & -- & --    \\ \hline
$2002$ & $2,520,504 $ & $26,838$  & \cellcolor{red!25}1,064.79 & 10/9 -- 10/21 & --  \\ \hline
$2003$ & $2,556,429 $ & $908$  &  \cellcolor{blue!25} 35.52 & -- & -- \\ \hline
$2004$ & $  2,631,831  $ & $154 $   &  \cellcolor{blue!25}  5.85 & -- &  -- \\ \hline
$2005$ & $2,673,560$ & $270 $   &  \cellcolor{blue!25}10.10 & 10/21 -- 10/31&  \\ \hline
$2006$ & $2,714,018$ & $ 377$   &  \cellcolor{blue!25}13.89 & -- & -- \\ \hline
$2007$ & $2,892,625 $ & $1,349$    &  \cellcolor{blue!25}46.64 & 10/6 -- 10/7 & 10/7 \\ \hline
$2008$ & $2,948,733$ & $2,476$ &  \cellcolor{blue!25}83.97 & -- & -- \\ \hline
$2009$ & $2,998,056$ & $6,819$  &  \cellcolor{red!25}227.45 &  6/9   & -- \\ 
       &             &          &  \cellcolor{red!25}       &  12/27   & \\ \hline
$2010$ & $2,675,656 $ & $6,159$    &  \cellcolor{red!25}230.19 & -- & -- \\ \hline
$2011$ & $2,693,606 $ & $5,321$  &  \cellcolor{red!25}197.54 & -- & --  \\ \hline
$2012$ & $2,710,968$ & $5,161$  &  \cellcolor{red!25} 190.37 & -- & -- \\ \hline
\end{tabular}
\begin{flushleft} 
\vspace{.5cm}
Incidence = Cases per $100,000$ inhabitants.  L.I stands for Linear Interpolation.
\end{flushleft}
\label{table7}
\end{table}

\begin{table}[t]
\centering
\caption{
{\bf S\~ao Lu\'is .}}
\begin{tabular}{|l|l|l|l|l|l|l|}
\hline
\multicolumn{1}{|l|}{\bf Year} & \multicolumn{1}{|l|}{\bf Pop.} & \multicolumn{1}{|l|}{\bf Cases}&\multicolumn{1}{|l|}{\bf Incidence}&\multicolumn{1}{|l|}{\bf Temp (CVX)} & \multicolumn{1}{|l|}{\bf Precip (L.I)} \\ \hline 
$2001$ & -- & --  &  --  & 10/1 -- 10/31  &    \\ 
       &  &   &    & 11/14   & --   \\ 
       & &   &    & 11/21  &    \\ \hline
$2002$ & $ 906,567$ & $448$  &  \cellcolor{blue!25}49.42 & 4/30 &   -- \\ \hline
$2003$ & $923,526 $ & $567$  &  \cellcolor{blue!25}61.40  &  9/5 -- 9/26 & -- \\ 
       &            &        &  \cellcolor{blue!25}       & 9/28 -- 10/10  & \\ \hline
$2004$ & $ 959,124  $ & $154 $   &  \cellcolor{blue!25}16.06 & -- & -- \\\hline
$2005$ & $978,824$ & $ 2,580$   &  \cellcolor{red!25}263.58 & -- & --   \\ \hline
$2006$ & $998,385$ & $1,395 $   &  \cellcolor{red!25}139.73 & -- & --  \\ \hline
$2007$ & $957,515 $ & $3,827$    &  \cellcolor{red!25}399.68  & -- & --  \\ \hline
$2008$ & $986,826$ & $1,183$ & \cellcolor{red!25}119.88  & -- & --  \\ \hline
$2009$ & $ 997,098$ & $100$  &  \cellcolor{blue!25}10.03  & -- & 5/31 \\ \hline
$2010$ & $ 1,014,837$ & $2,731$    &  \cellcolor{red!25}269.11 & -- &  -- \\ \hline
$2011$ & $1,027,430 $ & $5,229$  & \cellcolor{red!25}508.94  & 10/20 & --  \\ \hline
$2012$ & $1,039,610$ & $1,315$  & \cellcolor{red!25}126.49  &  6/8 -- 6/9   &     \\ 
       &             &          & \cellcolor{red!25}        & 6/12 -- 6/13   &  --    \\ 
       &             &          & \cellcolor{red!25}        &   7/24 -- 7/25  & \\ 
       &             &          & \cellcolor{red!25}        &   7/29  & \\ \hline
\end{tabular}
\begin{flushleft} 
\vspace{.5cm}
Incidence = Cases per $100,000$ inhabitants.  L.I stands for Linear Interpolation.
\end{flushleft}
\label{table8}
\end{table}

\end{document}